# A gas cloud on its way towards the super-massive black hole in the Galactic Centre

S.Gillessen[1], R.Genzel[1,2], T.K.Fritz[1], E.Quataert[3], C.Alig[4], A.Burkert[4,1], J.Cuadra[5], F.Eisenhauer[1], O.Pfuhl[1], K.Dodds-Eden[1], C.F.Gammie[6] & T.Ott[1]

[1] Max-Planck-Institut für extraterrestrische Physik (MPE), Giessenbachstr.1, D-85748 Garching, Germany ( ste@mpe.mpg.de, genzel@mpe.mpg.de )

[2] Department of Physics, Le Conte Hall, University of California, 94720 Berkeley, USA

[3] Department of Astronomy, University of California, 94720 Berkeley, USA

[4] Universitätssternwarte der Ludwig-Maximilians-Universität, Scheinerstr. 1, D-81679 München, Germany

[5] Departamento de Astronomía y Astrofísica, Pontificia Universidad Católica de Chile, Vicuña Mackenna 4860, 7820436 Macul, Santiago, Chile

[6] Center for Theoretical Astrophysics, Astronomy and Physics Departments, University of Illinois at Urbana-Champaign, 1002 West Green St., Urbana, IL 61801, USA

**Measurements of stellar orbits[1-3] provide compelling evidence[4,5] that the compact radio source Sagittarius A\* at the Galactic Centre is a black hole four million times the mass of the Sun. With the exception of modest X-ray and infrared flares[6,7], Sgr A\* is surprisingly faint, suggesting that the accretion rate and radiation efficiency near the event horizon are currently very low[3,8]. Here we report the presence of a dense gas cloud approximately three times the mass of Earth that is falling into the accretion zone of Sgr A\*. Our observations tightly constrain the cloud's orbit to be highly eccentric, with an innermost radius of approach of only ~3,100 times the event horizon that will be reached in 2013. Over the past three years the cloud has begun to disrupt, probably mainly through tidal**

**shearing arising from the black hole's gravitational force. The cloud's dynamic evolution and radiation in the next few years will probe the properties of the accretion flow and the feeding processes of the super-massive black hole. The kilo-electronvolt X-ray emission of SgrA\* may brighten significantly when the cloud reaches pericentre. There may also be a giant radiation flare several years from now if the cloud breaks up and its fragments feed gas into the central accretion zone.**

As part of our NACO[9] and SINFONI[10,11] Very Large Telescope (VLT) observation programmes studying the stellar orbits around the Galactic Centre super-massive black hole, Sgr A\*, we have discovered an object moving at about 1,700 km s$^{-1}$ along a trajectory almost straight towards Sgr A\* (Fig. 1). The object has a remarkably low temperature (about 550 K, Supplementary Fig. 2) and a luminosity about five times that of Sun, unlike any star we have so far seen near Sgr A\*. It is also seen in the spectroscopic data as a redshifted emission component in the Brγ and Brδ hydrogen lines, and the 2.058μm HeI line, with the same proper motion as the L'-band object. Its three-dimensional velocity increased from 1,200 km s$^{-1}$ in 2004 to 2,350 km s$^{-1}$ in 2011. The Brγ emission is elongated along its direction of motion with a spatially resolved velocity gradient (Fig. 2). Together these findings show that the object is a dusty, ionized gas cloud.

The extinction of the ionized gas is typical for the central parsec (Supplementary Information section 1) and its intrinsic Brγ luminosity is 1.66 (±0.25) × 10$^{-3}$ times that of the Sun. For case B recombination the implied electron density is $n_c = 2.6 \times 10^5 f_V^{-1/2} R_{c,15mas}^{-3/2} T_{e,1e4}^{0.54}$ cm$^{-3}$, for an effective cloud radius of R$_c$ ≈ 15 mas, volume filling factor $f_V$ (≤1) and an assumed electron temperature $T_e$ in units of $10^4$ K, a value typical for the temperatures measured in the central parsec[12]. The cloud mass is $M_c = 1.7 \times 10^{28} f_V^{1/2} R_{c,15mas}^{3/2} T_{e,1e4}^{0.54}$ g, or about $3 f_V^{1/2}$ Earth masses. It may plausibly be photo-ionized by the ultra-violet radiation field from nearby massive hot stars, as we infer from a comparison of the recombination rate with the number of impinging Lyman continuum photons[3,13]. This conclusion is supported by the HeI/Brγ line flux ratio of approximately 0.7, which is similar to the values found in the photo-ionized gas in the central parsec (0.35–0.7). If so, the requirement of complete photo-ionization sets a lower limit to $f_V$ of $10^{-1 \pm 0.5}$ for the extreme case that the cloud is a thin sheet.

The combined astrometric and radial velocity data tightly constrain the cloud's motion. It is on a highly eccentric (e = 0.94) Keplerian orbit bound to the black hole (Fig. 1, Table 1, Supplementary Information section 2). The pericentre radius is a mere 36 light hours (3,100 Schwarzschild radii, $R_S$), which the cloud will reach in summer 2013. Only the two stars S2 ($r_{peri}$ = 17 light hours) and S14 ($r_{peri}$ = 11 light hours) have come closer to the black hole[2,3] since our monitoring started in 1992. Although the cloud's gas density may be only modestly greater than other ionized gas clouds in the central parsec – $n_e \approx (0.1-2) \times 10^5$ cm$^{-3}$; refs 12 and 14 – it has a specific angular momentum about 50 times smaller[12].

For the nominal properties of the X-ray detected accretion flow onto the black hole[15,16] the cloud should stay close to Keplerian motion all the way to the pericentre (Supplementary Information sections 3 and 4). Its density currently is about $300 f_V^{-1/2}$ times greater than that of the surrounding hot gas in the accretion flow[15]; extrapolating to pericentre its density contrast will then still be about $60 f_V^{-1/2}$. Similarly, the cloud's ram pressure by far exceeds that of the hot gas throughout the orbit. In contrast, the thermal pressure ratio will quickly decrease from unity at apocentre and the hot gas is expected to drive a shock slowly compressing the cloud. Whereas the external pressure compresses the cloud from all directions, the black hole's tidal forces shear the cloud along the direction of its motion, because the Roche density for self-gravitational stabilization exceeds the cloud density by nine orders of magnitude[3]. In addition, the ram pressure compresses the cloud parallel to its motion. The interaction between the fast-moving cloud and the surrounding hot gas should also lead to shredding and disruption, owing to the Kelvin-Helmholtz and Rayleigh-Taylor instabilities[17-20]. Rayleigh-Taylor instabilities at the leading edge should in fact break up the cloud within the next few years if it started as a spheroidal, thick blob (Supplementary Information section 3). A thin, dense sheet would by now already have fragmented and disintegrated, suggesting that $f_V$ is of the order of unity.

We are witnessing the cloud's disruption happening in our spectroscopic data (Fig. 2). The intrinsic velocity width more than tripled over the last eight years, and we see between 2008 and 2011 a growing velocity gradient along the orbital direction. Test particle calculations implementing only the black hole's force show that an initially spherical gas cloud placed on the orbit (Table 1) is stretched along the orbit and compressed perpendicular to it, with increasing velocity widths and velocity gradients reasonably matching our observations (Fig. 3, Supplementary Fig. 4). There is also a tail of gas with lower surface brightness on approximately the same orbit as the cloud, which cannot be due to tidal disruption alone. It may be stripped gas, or lower-density, lower-filling-factor gas on the same orbit. The latter explanation is more plausible given that the integrated Brγ and L'-band luminosities did not drop by more than 30% between 2004 and 2011, and the integrated Brγ flux of the tail is comparable to that of the cloud.

The disruption and energy deposition processes in the next years until and after pericentre are powerful probes of the physical conditions in the accretion zone (Supplementary section 3). We expect that the interaction between hot gas and cloud will drive a strong shock into the cloud. Given the densities of cloud and hot gas, the cloud as a whole should remain at low temperature until just before it reaches the pericentre. Near the pericentre the post-shock temperature may increase rapidly to $T_{postshock,\ c} \sim 6-10 \times 10^6$ K, resulting in X-ray emission. We estimate the observable 2−8keV luminosity to be less than $10^{34}$ erg s$^{-1}$ there, somewhat larger than the current 'quiescent' X-ray luminosity[6,15,21] of Sgr A* ($10^{33}$ erg s$^{-1}$). Rayleigh-Taylor instabilities may by then have broken up the cloud into several sub-fragments, in which case the emission may be variable throughout this period. Our predictions depend sensitively on the density and disruption state of the cloud, as well as on the radial dependencies of the hot gas properties, none of which we can fully quantify. The steeper the radial profiles are and the higher the value of $f_V$, the more X-ray emission will occur. Shallower profiles and a low value of $f_V$ could shift the emission into the un-observable soft X-ray and ultraviolet bands. Together the evolution of the 2-8keV and Brγ luminosities, as well as the Brγ velocity distribution will strongly constrain the thermal states and interaction of the cloud and the ambient hot gas in the at present un-probed regime of $10^3$ R$_S$ − $10^4$ R$_S$, when compared with test particle calculations and more detailed numerical simulations (Fig. 3 and Supplementary Fig. 4).

The radiated energy estimated above is less than 1% of the total kinetic energy of the cloud, about $10^{45.4}$ erg. As the tidally disrupted filamentary cloud passes near pericentre some fraction of the gas may well collide with itself, dissipate and circularize[22]. This is probable because of the large velocity dispersion of the cloud, its size comparable to the impact parameter and because the Rayleigh-Taylor and Kelvin-Helmholtz time scales are similar to the orbital time scale. Because the mass of the cloud is larger than the mass of hot gas within the central 3,100 $R_S$ or so (approximately $10^{27.3}$ g; ref. 15), it is plausible that then the accretion near the event horizon will be temporarily dominated by accretion of the cloud. This could in principle release up to around $10^{48}$ erg over the next decade, although the radiative efficiency of the inflow at these accretion rates is of order 1-10% (refs 23 and 24). Observations of the emission across the electromagnetic spectrum during this post-circularization phase will provide stringent constraints on the physics of black-hole accretion with unusually good knowledge of the mass available.

What was the origin of the low-angular-momentum cloud? Its orbital angular momentum vector is within 15° of the so-called 'clock-wise' disk of young, massive O and Wolf-Rayet stars at radii of about 1" to 10" from Sgr A* (refs 3 and 25). Several of these stars have powerful winds. One star, IRS16SW, about 1.4" southeast of Sgr A* is a massive, Wolf-Rayet contact binary[26]. Colliding winds in the stellar disk, and especially in binaries, may create low angular momentum gas that then falls deep into the potential of the supermassive black hole[27,28].

Supplementary Information is linked to the online version of the paper at www.nature.com/nature.

Acknowledgments*:* This paper is based on observations at the Very Large Telescope (VLT) of the European Observatory (ESO) in Chile. We thank Chris McKee and Richard Klein for helpful discussions on the cloud destruction process. J.C. acknowledges support from FONDAP, FONDECYT, Basal and VRI-PUC.

Author contributions: S.G. collected and analyzed the data and discovered the orbit of the gas cloud. R.G. and S.G. wrote the paper. T.F. detected the high proper motion and extracted the astrometric positions and the photometry. R.G., A.B. and E.Q. derived the cloud's properties, its evolution and the estimate of the X-ray luminosity. R.G., E.Q., A.B. and C.G. contributed to the analytical estimates. C.A. and J.C. set up numerical simulations to check the analysis. F.E., O.P. and K.D.-E. helped in the data analysis and interpretation. T.O. provided valuable software tools.

Author information: The authors have no competing financial interests.

## Table 1. Orbit Parameters of the Infalling Cloud

| parameters of Keplerian orbit around $4.31 \times 10^6$ M$_\odot$ black hole at R$_0$ = 8.33 kpc | best fitting value |
|---|---|
| semi-major axis a | 521 ± 28 milli-arcsec |
| eccentricity e | 0.9384 ± 0.0066 |
| inclination of ascending node i | 106.55 ± 0.88 degrees |
| position angle of ascending node Ω | 101.5 ± 1.1 degrees |
| longitude of pericentre ω | 109.59 ± 0.78 degrees |
| time of pericentre t$_{peri}$ | 2013.51 ± 0.035 |
| pericentre distance from black hole r$_{peri}$ | $4.0 \pm 0.3 \times 10^{15}$ cm = 3140 R$_S$ |
| orbital period t$_o$ | 137 ± 11 years |

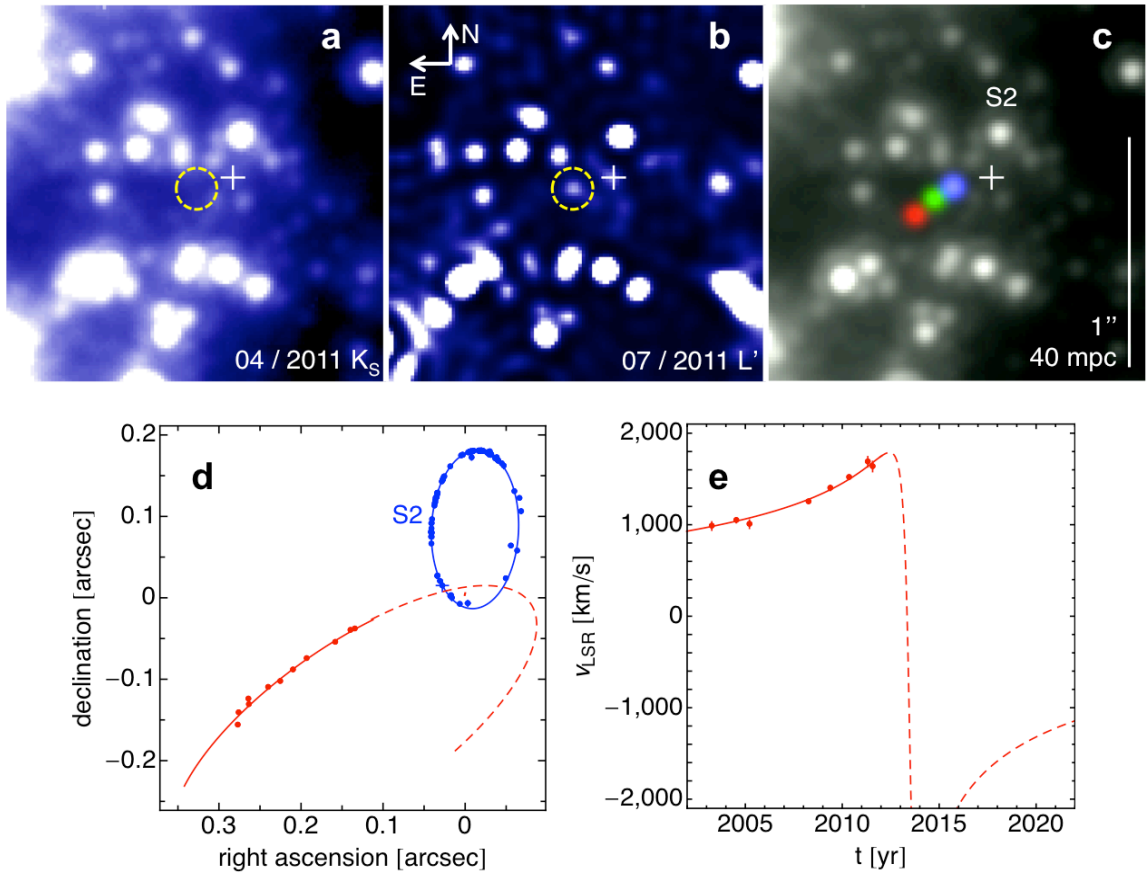

**Figure 1: Infalling dust/gas cloud in the Galactic Centre. a** and **b**, NACO[9] adaptive optics VLT images showing that the cloud (dashed circle) is detected in the L'-band (3.76 μm) but not in $K_s$-band (2.16 μm), indicating that it is not a star but a dusty cloud with a temperature of around 550 K (Supplementary Fig. 2). The cloud is also detected in the M-band (4.7μm) but not seen in the H-band (1.65 μm). North is up, East is left. The white cross marks the position of Sgr A*. **c**, The proper motion derived from the L-band data is about 42 mas yr$^{-1}$, or 1,670 km s$^{-1}$ (in 2011), from the southeast towards the position of Sgr A* (red for epoch 2004.5, green for 2008.3 and blue for 2011.3, overlaid on a 2011 $K_s$-band image). The cloud is also detected in deep spectroscopy with the adaptive optics assisted integral field unit SINFONI[10,11] in the HI n=7-4 Brγ recombination line at 2.1661 μm and in HeI at 2.058 μm, with a radial velocity of 1,250 km/s (in 2008) and 1,650 km/s (in 2011). **d** and **e**, The combination of the astrometric data in L' and Brγ and the radial velocity ($v_{LSR}$) data in Brγ tightly constrains the orbit of the cloud (error bars are 1σ measurement errors). The cloud is on a highly eccentric, bound orbit (e = 0.94), with a pericentre radius and time of 36 light hours (3,100 $R_S$) and 2013.5 (Table 1). For further details see the Supplementary Information.

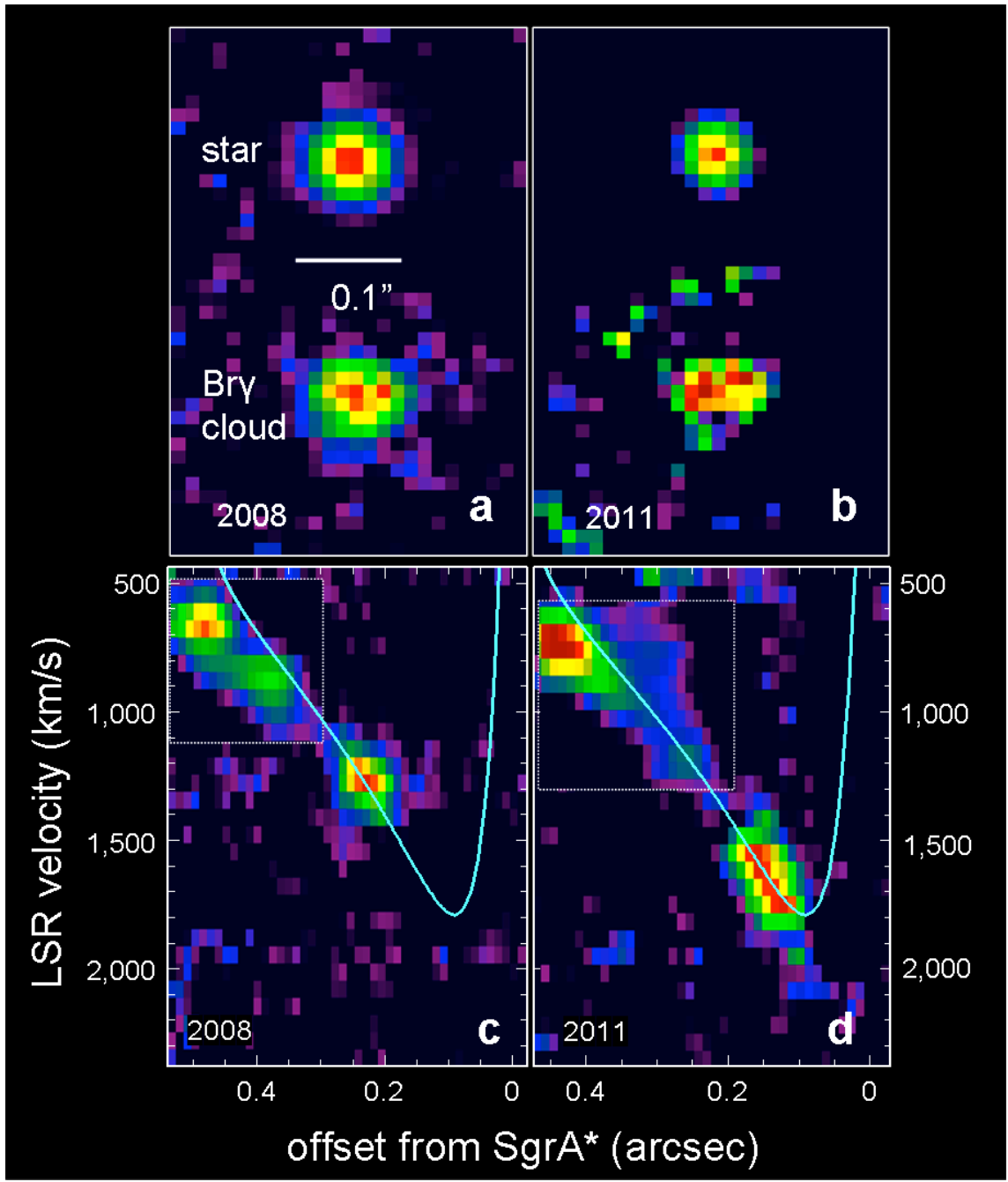

**Figure 2: The velocity shear in the gas cloud**. The left column shows data from 2008.3, the right from 2011.3. Panels **a** and **b** show integrated Brγ maps of the cloud, in comparison to the point spread function from stellar images shown above. The inferred intrinsic East-West half-width at half-maximum source radii are $R_c$ = 21 ± 5 mas in 2008 and 19 ± 8 mas in 2011 (approximately along the direction of orbital motion), after removal of the instrumental broadening

estimated from the stellar images above. A similar spatial extent is found from the spatial separation between the red- und blue-shifted emission of the cloud ($R_c$ = 23 ± 5 mas). The minor-axis radius of the cloud is only marginally resolved or unresolved (radius less than 12 mas). We adopt $R_c$ = 15 mas as the 'effective' circular radius, from combining the results in the two directions. Panels **c** and **d** are position-velocity maps, obtained with SINFONI on the VLT, of the cloud's Brγ emission. The slit is oriented approximately along the long axis of the cloud and the projected orbital direction and has a width of 62 mas for the bright 'head' of the emission. For the lower surface brightness 'tail' of emission (in the enclosed white dotted regions) we smoothed the data with 50 mas and 138 km s$^{-1}$ and used a slit width of 0.11". The gas in the tail is spread over around 200 mas downstream of the cloud. The trailing emission appears to be connected by a smooth velocity gradient (of about 2km s$^{-1}$ mas$^{-1}$), and the velocity field in the tail approximately follows the best-fit orbit of the head (cyan curves, see also Table 1). An increasing velocity gradient has formed in the head between 2008 (2.1 km s$^{-1}$ mas$^{-1}$) and 2011 (4.6 km s$^{-1}$ mas$^{-1}$). As a result of this velocity gradient, the intrinsic integrated full-width at half-maximum (FWHM) velocity width of the cloud increased from 89 (± 30) km s$^{-1}$ in 2003 and 117 (± 25) km s$^{-1}$ in 2004, to 210 (± 24) km s$^{-1}$ in 2008, and 350 (± 40) km s$^{-1}$ in 2011.

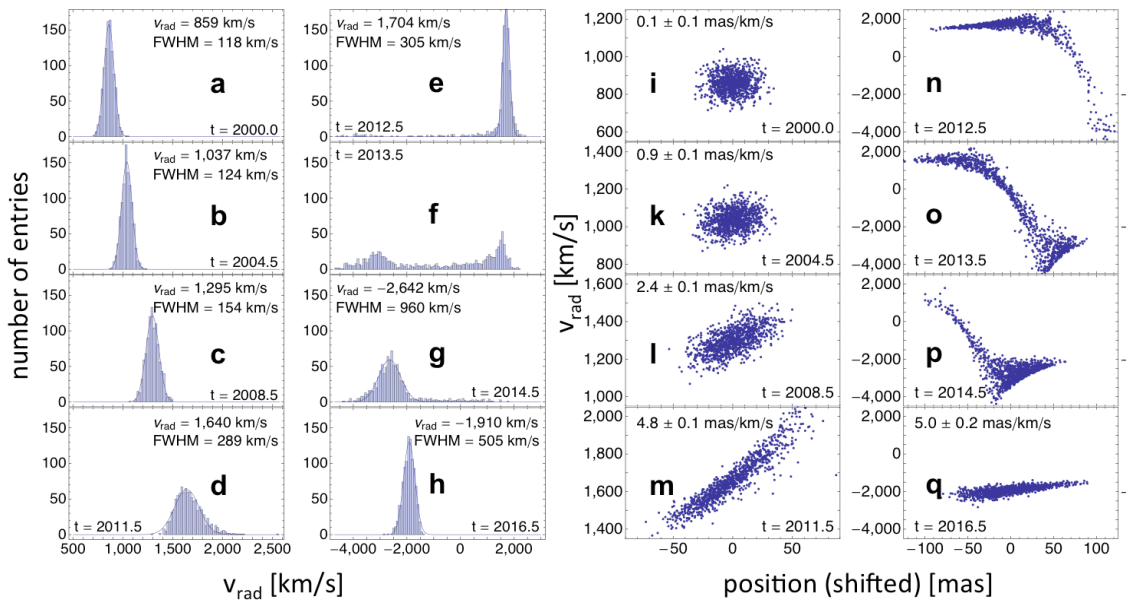

**Figure 3. Test particle simulation of the orbital tidal disruption.** An initially Gaussian cloud of initial FWHM diameter 25 mas and FWHM velocity width 120 km s$^{-1}$ is placed on the orbit in Table 1. **a–h**, As described in Supplementary Information section 4 (see also Supplementary Fig. 4) these panels show the evolution of the cloud integrated velocity width for eight epochs. The mean velocity and FWHM are given for those epochs at which the distribution is unimodal (that is, not **f**). **i–p,** These panels show the evolution of the velocity change as a function of position (measured in milli-arcseconds) along the orbital direction, purely on the basis of the tidal disruption of the cloud by the gravitational force of the super-massive black hole. This toy model is a good description of the velocity (and spatial) data between 2004 and 2011, thus allowing plausible forward projections until pericenter passage. Beyond that, the test particle approach will probably fail owing to the hydrodynamic effects, which will then most probably dominate further evolution.

# Supplementary Information

## S1. Description of observations and data analysis

Our results are based on observations at the Very Large Telescope (VLT) of the European Observatory (ESO) in Chile. We adopt a Sun − Galactic Centre distance of 8.33 kpc, where 1" = 40 mpc = $1.25 \times 10^{17}$ cm, or $9.8 \times 10^4$ times the Schwarzschild radius ($R_S$) of the super-massive black hole ($M_{BH} = 4.31 \times 10^6$ $M_\odot$)[1-3].

*Astrometry*

We have been monitoring the central few arcseconds of the Milky Way since 2002 with the adaptive optics imager NACO[9] at the VLT. By means of its infrared wavefront sensor, this instrument allows the use of a natural guide star even in the Galactic Centre field, where the high extinction ($A_V$>30 mag) would otherwise profit from using a laser guide star. We used the bright ($m_K = 6.5 − 7.0$) supergiant IRS7 located roughly 5.5" North of SgrA* as a guide star. L'-band (3.76μm) data were obtained as part of various scientific programmes, such as studying the gaseous structures in the Galactic Centre, or the activity of SgrA* itself. Also, L'-band data were obtained when the atmospheric conditions prevented us from using shorter wavelengths. Mostly, a random jitter mode was applied since a median of a sufficiently large number of such frames in L'-band is as good an estimate for the thermal background as dedicated sky frames. The sampling is 27 mas/pix, well-matched to the resolution in L'-band. We reduced the data following the standard techniques, including sky-subtraction, flat-fielding and bad pixel correction. After a quality selection, we registered the individual frames and created the final maps. This yielded essentially one map for each year since 2002. We used the analysis tool 'starfinder' [29] for extraction of the point spread function for each map. In order to measure positions on the maps, we followed our usual technique[2] of Lucy deconvolution[30]. The central regions of the final maps, following deconvolution and restoration with a Gaussian kernel of FWHM 2 pixels (54 mas), are shown in Figure S1. In these, we measured the positions of the object marked by a yellow arrow in Figure S1 and neighbouring stars with Gaussian fits. The conversion to astrometric coordinates is achieved by defining a local coordinate system by means of the stars, for which astrometric coordinates are known[2]. We estimate the accuracy of the relative astrometry to be around 2 mas, dominated by the complicated background, mostly due to the seeing halos of the surrounding sources.

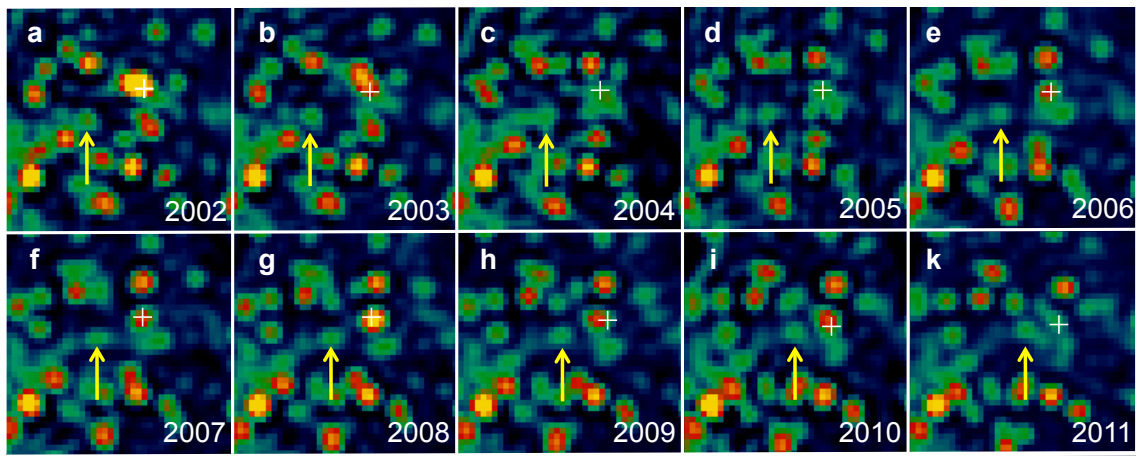

Supplementary Figure S1. Final L'-band maps, after Lucy deconvolution and restoration with a Gaussian kernel of FWHM 54 mas, centered on the object (marked by the yellow arrows). North is up, East is left. The box size is 1". The white cross marks the position of Sgr A* which, in many of the frames, is flaring.

*Spectroscopy*

The radial velocity information originates from observations with the adaptive optics assisted integral field spectrometer SINFONI[10,11] obtained between 2004 and 2011, plus one data point from 2003, when the integral field unit of SINFONI was operated at the VLT without adaptive optics as a guest instrument. SINFONI does not offer infrared wavefront sensing and hence we used a fainter optical guide star further away. It has a magnitude of $m_R$ = 14.65 and is located 10.8" East and 18.8" North of SgrA*. As a consequence, the data quality depends strongly on the atmospheric conditions. Data reduction was as follows: sky frames were subtracted first, followed by calibration of the detector artefacts (flat-fielding and bad pixel removal), and finally data cubes were constructed including a distortion correction and wavelength calibration from emission line lamps and the atmospheric OH-lines. Finally, all cubes per epoch were co-added, which also strongly suppresses cosmic ray hits. We detected the object in Brγ line emission redshifted by 1000 km/s to 1700 km/s in eight of our cubes. Most of these cubes cover the H- and K-band simultaneously with a spectral resolution of R=1500, some are K-band only with R=4000. Given the crowding in the field, only the cubes obtained under good conditions with the smallest pixel scale (12.5 mas/pix) are useful, except for the 2003 data when no such data were taken and we were able to see the same line in a cube with 50 mas/pix. The three deepest and highest quality cubes are from 2004.53, 2008.27 and 2011.3 with total on-source integration times of around 4 hours each. No deconvolution was applied.

*Photometry and nature of the fast moving object*

Since we have NACO images at different wavebands, we can constrain the spectral energy distribution of the object. Of interest here are $K_s$-band (2.16μm), L'-band (3.76μm) and M-band (4.7μm). In the L'-band, we use the same data as for the astrometry to obtain an average flux density (within the measurement uncertainties, we do not see any variability). In the few M-band data sets the object is detected as well. We use the highest quality data set from September 20, 2004. In $K_s$-band, the object is not detectable in the data sets we use for the astrometry of the stellar orbits. Using high quality data sets, we infer that the object needs to be fainter than $m_{Ks}$ = 17.8 using a 3σ threshold and a minimum correlation coefficient of 0.7 in 'starfinder'. The same tool is used for determining the object's flux densities, which we calibrate with the two bright stars IRS16C ($m_{Ks}$ = 9.93, $m_{L'}$ = 20)[31] and IRS16NW ($m_{Ks}$ = 10.14, $m_{L'}$ = 8.43)[31]. In the M-band we assume the same absolute magnitudes for the calibrators as in L'-band. The extinction values adopted are those from [32] for the calibrators, for the object we use color corrected values of $A_{Ks}$ = 2.22, $A_{L'}$ = 1.07, $A_M$ = 0.94. The resulting photometry is shown in Figure S2. The best-fitting blackbody temperature is 550 (+140/-90) K. The $K_s$-band limit implies T < 640 K. We conclude that the dust in the object has a temperature of $550 \pm 90$ K. The fairly high dust temperature suggests that the continuum emission comes from small (~ 20 nm), transiently heated dust grains[33] with a total warm dust mass of $2 \times 10^{23}$ g, a small fraction of the cloud's mass, although there may be additional colder dust.

Given the low temperature and simultaneous relatively bright HI and HeI emission, there is no known star with such properties. The most plausible source of the faint (L~5 $L_\odot$), cool infrared emission is warm dust. The inferred dust temperature is somewhat cooler than but similar to the dusty clumps seen in the dusty HII region around the compact stellar group IRS13E, seen 3.5" South-West in projection from Sgr A*[34]. The size of the gas cloud would be 1.2 (+1.1/-0.5) AU if it were an optically thick blackbody. This would be roughly the size of a giant star and is much smaller than the actual size measured in Brγ. Hence, the object must be optically thin to infrared radiation. Also, the extinction appears to be normal at the position of the object, so it is unlikely that any embedded source is hidden behind local extinction. The ratio of Brγ and Brδ is as expected for a typical value of the extinction toward SgrA*[32].

The discussion in the last paragraph shows that any central stellar source embedded in the dust/gas cloud must be sufficiently hot (>$10^{4.6}$ K, so that it emits mostly in the ultraviolet) and at the same time have a low enough luminosity (<$10^{3.7}$ $L_\odot$, so that its 2μm emission in the Rayleigh-Jeans tail is below our detection limit). The only stellar sources that could match the emission lines and the K-band limit are compact planetary nebulae[35,36]. They have luminosities of order ~$10^{3.8} L_\odot$, and even have a sufficient number of ionizing photons ($Q_{Lyc}$~$10^{46.7}$ $s^{-1}$) to fully ionize the cloud from within. At the same time the central stars are hot enough (> $10^{4.5...4.9}$ K) that they are below our detection limit in the K-band or at longer wavelengths ($m_{Ks}$>17…19). The L- and M-band emission would then come from dust in the circum-stellar shell that is also seen in the HeI and Brγ/Brδ lines. Depending on the age of the system, the typical gas shell radii of planetary nebulae are 0.03 to 0.3 pc[37]. For this size and for a ~0.6 $M_\odot$ central star, most of the planetary nebula thus cannot survive anywhere in the central parsec because of the tidal forces from the super-massive black hole and nuclear star cluster, as well as additional hydrodynamic interactions with stellar winds and the ionized gas. However, for a young and compact (proto)-planetary nebula on a highly eccentric orbit the innermost circum-stellar gas ($R<10^{15}$ cm) may still survive to r~1". Inside of this radius even this compact circum-stellar gas would be tidally disrupted. Its dynamical evolution would then be indistinguishable from a gas cloud without a central star. Hence, the possible presence of such a low mass star is interesting in terms of the origin of the cloud but otherwise has no impact on the discussion on the cloud dynamics and evolution.

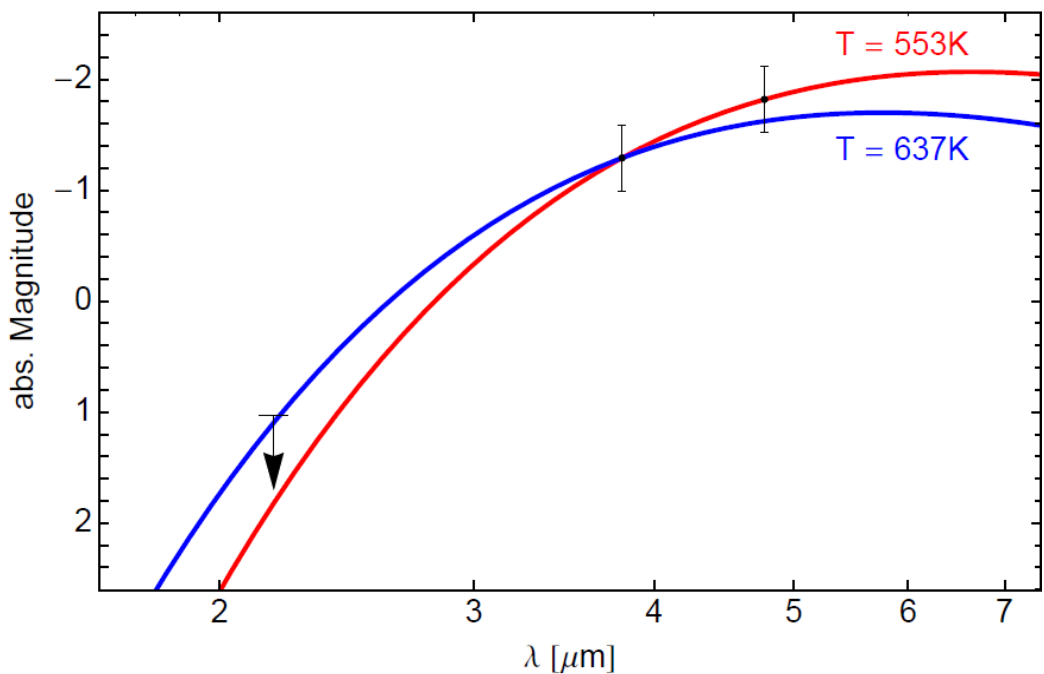

Supplementary Figure S2. Spectral energy distribution of the object and inferred temperatures of equivalent black bodies. The blue line corresponds to T = 637 K. This is the hottest temperature, which does not violate the K-band limit given the L-band flux. The red line is for 553 K, the best matching temperature given the L- and M-band fluxes. We conclude T = 550 (±90) K. The integrated luminosity is ~ 5 $L_\odot$

## S2. Orbit fitting

We used the same tools as in [2] for fitting the orbit. The gravitational potential assumed was also taken from that work. The orbit of the gas cloud is well defined. The astrometric acceleration is significant at the 10σ-level. Furthermore, the measured radial velocity of the cloud is changing (Figure 1). Hence, seven dynamical quantities are known (R.A., Dec., $v_{R.A.}$, $v_{Dec.}$, $v_{rad}$, $a_{2D}$, $a_{rad}$) and it is non-trivial that they can be described by one orbit (which has six degrees of freedom). Figure S3 shows the parameter uncertainties for five of the parameters from a Markov-Chain Monte Carlo analysis. The orbital elements are given in Table 1. The orientation of the orbit coincides with the plane of the inner edge of the stellar disk of young stars[26].

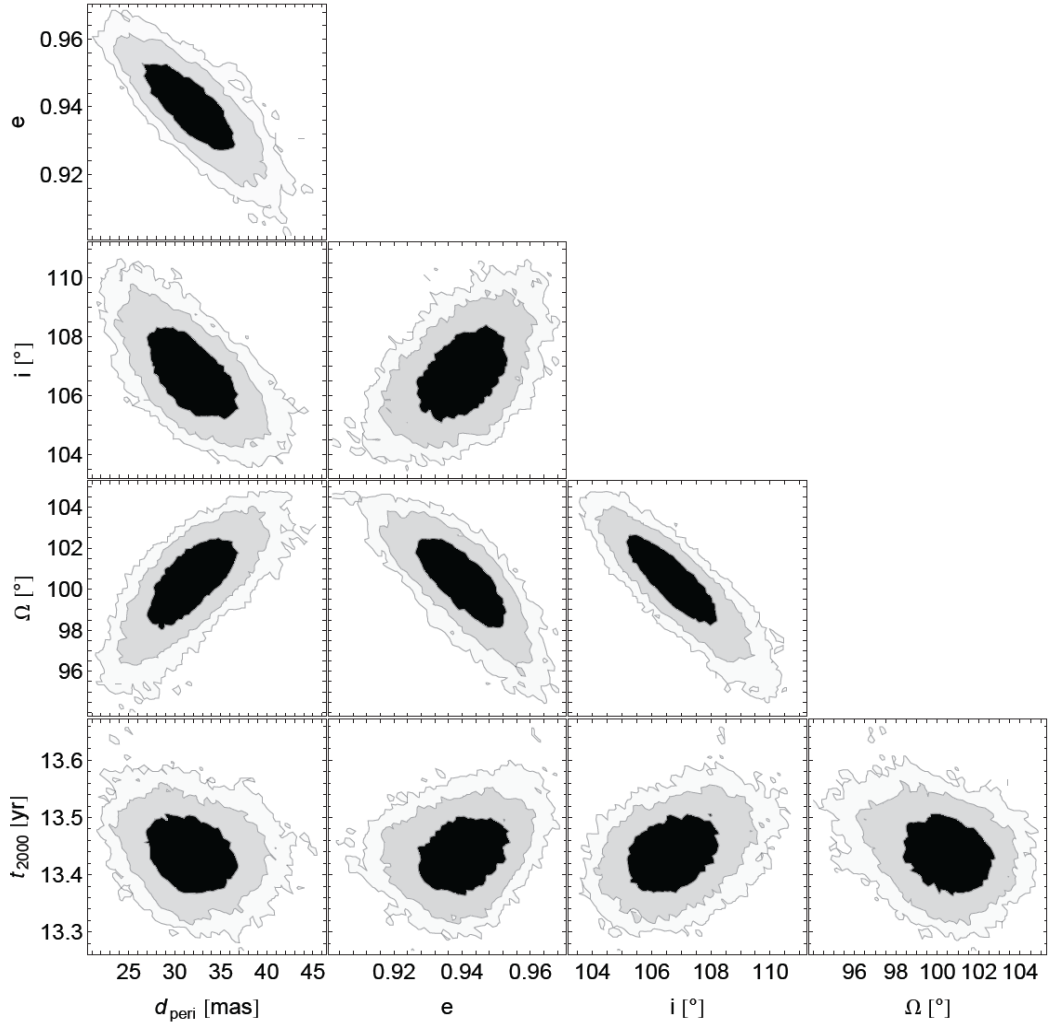

Supplementary Figure S3. Parameter uncertainties for five of the six orbital elements (pericentre distance, eccentricity, inclination, position angle of ascending node, and time of pericentre passage) from Markov chain simulations. The contours delineate the 1σ, 2σ and 3σ levels.

## S3: Hydrodynamics of the infall

Models of the X-ray emission from Sgr A* yield the density and temperature of the hot gas in the accretion zone, within the Bondi radius of the super-massive black hole ($r_B \sim 10^5 R_S \sim 1$"), to be[15,16]

$$n_{hot} \sim 930 \left(\frac{1.4 \times 10^4 R_S}{r}\right)^{\alpha} \text{ (cm}^{-3}\text{)}, \quad T_{hot} \sim 2.1 \times 10^8 \left(\frac{1.4 \times 10^4 R_S}{r}\right)^{\beta} \text{ (K)} \quad \text{(eS1)},$$

where $\alpha \sim \beta \sim 1$. At $r_{2011} \sim 1.4 \times 10^4 R_S$ the cloud has a density that is $\sim 300 f_V^{-1/2}$ times greater, and a ram pressure ($3.3 \times 10^{-2} f_V^{-1/2}$ erg cm$^{-3}$) that is $600 f_V^{-1/2}$ times greater than those of the surrounding hot gas, where $f_V$ is the volume filling factor of the observed Brγ emission. If we extrapolate the model parameters to pericentre ($r_p=3100 R_S$) the cloud to hot gas density contrast and ram pressure ratio should then still be $\sim 60 f_V^{-1/2}$ and $150 f_V^{-1/2}$, respectively. We thus expect that the cloud stays close to Keplerian motion all the way to pericentre.

For a constant cloud density or filling factor the cloud's thermal pressure ($\sim 3.6 \times 10^{-7} f_V^{-1/2} T_{e,1e4}^{1.54}$ erg cm$^{-3}$, with $T_{e,1e4}$ the electron temperature in units of $10^4$ K) falls rapidly below that of the hot gas as the cloud approaches the black hole, from pressure equilibrium at apocentre ($r_{apo} \sim r_B \sim 1$"). Given the orbital parameters, and for $f_V \sim 1$ and the hot gas parameters in equation (eS1), we expect that the interaction between hot gas and cloud drives a strong shock into the cloud. It compresses the cloud from all directions on a time scale

$$t_{cp} = \frac{R_c}{c_{hot}} \times \left(\frac{n_c}{n_{hot}}\right)^{1/2} = 11 R_{c,15mas} \times \left(\frac{r}{1.4 \times 10^4 R_S}\right) \quad \text{(yr)} \quad \text{(eS2)},$$

where $R_c$ is the cloud radius, $n_c$ and $n_{hot}$ are the volume densities of the cloud and hot gas in the accretion flow, and $c_{hot}$ is the sound speed in the hot gas. We find that $t_{cp}$ is $1 - 5$ times the dynamical time scale along the orbit and 0.2 times the sound crossing time of the cloud. The cloud thus will develop a growing very dense shell surrounding an inner zone at the original density. The fragmentation and shredding of the cloud at its surface, due to the Kelvin-Helmholtz and Rayleigh Taylor instabilities, happens on similar time scales of[17-20]

$$t_{KH} \sim 4 \times R_{c,15mas} \left(\frac{n_c}{2.6 \times 10^5 \text{ cm}^{-3}}\right)^{1/2} \left(\frac{r}{1.4 \times 10^4 R_S}\right) \text{ (yr), and}$$

$$t_{RT} \sim 4 \times R_{c,15mas}^{1/2} \times \left(\frac{r}{1.4 \times 10^4 R_S}\right) \times f_V \qquad \text{(yr)} \quad \text{(eS3)}.$$

These estimates indicate that as the cloud approaches pericentre, it is getting denser but at the same time it inevitably fragments via the Rayleigh-Taylor instability, and may also get shredded on its surface via the Kelvin-Helmholtz instability. The post-shock temperature is

$$T_{postshock,c} \sim 2.6 \times 10^5 f_V^{1/2} R_{15mas}^{3/2} T_{e,1e4}^{-0.54} \left(\frac{r}{1.4 \times 10^4 R_S}\right)^{-2} \text{ (K)} \quad \text{(eS4)}.$$

The cloud with an initial post-shock density $n_{c,ps} > 4n_c$ (for an adiabatic index $\gamma = 5/3$) cools on a time

$$t_{cool} < 3.4 \times 10^{-3} f_V^{0.85} R_{15mas}^{2.5} T_{e,1e4}^{-0.91} \left(\frac{n_{c,ps}}{10^6 cm^{-3}}\right) \left(\frac{r}{1.4 \times 10^4 R_S}\right)^{-3.4} \text{ (yr)} \quad \text{(eS5)},$$

where we used cooling curves[38] at twice solar metallicity, characteristic of the gas abundances in the central parsec[3]. The ratio of cooling time to dynamical time scale is much less than unity throughout the orbit until early 2013, at which point it increases to a value of ~3 at pericentre. With the nominal parameters of the accretion flow in equation (eS1), the cloud as a whole should thus remain at low temperature until just before pericentre. Near pericentre the post-shock temperature may increase rapidly to $T_{pc} \sim 6 - 10 \times 10^6$ K with a total luminosity of

$$L_{cool} = \frac{M_c}{\mu}\frac{3}{2}kT_{pc}/t_{cool} = 10^{35.6} f_V^{0.25} R_{15mas}^{0.5} \left(\frac{n_{c,postshock}}{10^6 cm^{-3}}\right) \left(\frac{r}{3100\, R_S}\right)^{1.4} \text{ (erg/s)} \quad \text{(eS6)}.$$

For a Galactic foreground absorption column of $N(H)_{abs} \sim 4-9 \times 10^{22}$ cm$^{-2}$ [39] the observable $2 - 8$ keV luminosity derived from equation (eS6) is $\leq 10^{34}$ erg/s for the parameters near pericentre, somewhat larger than the current 'quiescent' X-ray luminosity of Sgr A* ($10^{33}$ erg/s)[15, 21].

The sensitivity of the first order predictions in equations (eS4)-(eS6) above on the radial dependencies of the hot gas properties can be understood in terms of postshock temperature and cooling time. If the radial profiles are significantly steeper than assumed ($\alpha, \beta > 1$) the postshock temperature and cooling time increase, leading to more 2-8 keV X-ray emission. In turn, if the density and temperature profiles are shallower, the cooling is more efficient but most of it occurs at shorter wavelengths and might be un-observable. A low cloud volume filling factor combined with higher density would yield similar observational signatures, i.e. little or no X-ray emission due to the low temperature. This degeneracy, however, will be broken after the pericentre passage. Comparing the post-pericentre Brγ data with the test particle simulations will thus constrain $f_V$. This in turn can then be used to interpret the evolution of the X-ray luminosity during the pericentre passage, the free parameters being the radial temperature and density profiles of the hot gas.

The evolution of the cloud after pericentre passage probably is dynamically chaotic, unless the cloud consists of many small sub-clumps with higher local density. For $f_V \approx 1$, which we favour because of the photoionization of the cloud, the Kelvin-Helmholtz and Rayleigh Taylor time scales are comparable to the dynamical time scale, the cloud's velocity dispersion is large, and the impact parameter is similar to the cloud's extension. This means that most probably the cloud can efficiently circularize, in agreement with what we find in preliminary numerical gas simulations.

## S4. Test particle simulations of tidal disruption

We have simulated the time evolution of the size, velocity width and velocity gradients from a test particle calculation of a gas cloud in the gravitational potential of the super-massive black hole in the Galactic Centre (without the effects of the hot gas in the accretion flow). The model consists of 1000 test particles, distributed in a Gaussian fashion in phase space, with a FWHM in space coordinates of 25 mas and 120 km/s in velocity coordinates. We chose the epoch 2000.0 for these initial conditions. The orbits of the particle 'cloud' in the gravitational potential of the super-massive black hole yields the evolution in size, velocity width and velocity shear shown Figures 3 and S4. They resemble our observations relatively closely. We have varied the initial size and velocity width of the cloud, as well as the density distribution (constant vs. Gaussian), which shows that the results are not strongly dependent on these initial conditions. The cloud starts to spread significantly around 2008, mainly in the spatial direction along the orbit and in radial velocity. It also develops a velocity gradient similar to the observed one. During the passage, the cloud gets tidally stretched and then comes back to a more symmetric, albeit wider, configuration. At the same time, the tidal effects compress the cloud perpendicular to its motion.

We don't expect that the cloud's evolution will follow this simple test particle calculation, given the complex dynamical processes at play beyond gravitational forces. Nevertheless, it is instructive to compare the actual evolution with the test particle case, which essentially represents the evolution if the accretion flow would be absent.

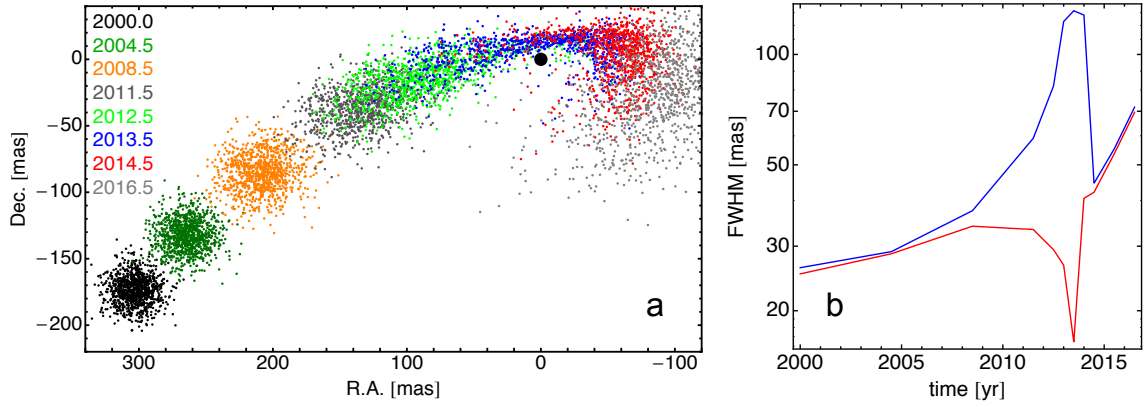

Supplementary Figure S4. Simulation of orbital tidal disruption of Gaussian cloud: evolution of spatial structure as a function of time (panel a, North is up, East is left). Panel b shows the evolution of the FWHM along directions parallel (blue) and perpendicular (red) to the orbit.

**Additional References for Supplementary Information**